\documentclass[%
 aip,
 sd,%
 amsmath,amssymb,
 reprint,%
]{revtex4-1}

\usepackage{ulem} 

\usepackage{graphicx}
\usepackage{dcolumn}
\usepackage{bm}

\begin{document}


\title{Enhanced power factor and reduced Lorenz number in the Wiedemann--Franz law due to pudding mold type band structures} 



\author{Hidetomo Usui}
\affiliation{Department of Physics, Osaka University, Machikaneyama-cho, Toyonaka, Osaka, 560-0043, Japan}
\author{Kazuhiko Kuroki}
\affiliation{Department of Physics, Osaka University, Machikaneyama-cho, Toyonaka, Osaka, 560-0043, Japan}


\date{\today}

\begin{abstract}
We study the relationship between the shape of the electronic band structure and the thermoelectric properties.
In order to study the band shape dependence of the thermoelectric properties generally, we first adopt models with band structures having the dispersion $E({\bf k}) \sim |{\bf k}|^n$ with $n = 2, 4$ and 6. We consider one, two- and three dimensional systems, and calculate the thermoelectric properties using the Boltzmann equation approach within the constant quasi-particle lifetime approximation.
$n = 2$ corresponds to the usual parabolic band structure, while the band shape for $n = 4, 6$ has a flat portion at the band edge, so that the density of states diverges at the bottom of the band.
We call this kind of band structure the ``pudding mold type band".  $n \ge 4$ belong to the pudding mold type band, but since the density of states diverges even for $n= 2$ in one dimensional system, this is also categorized as the pudding mold type. Due to the large density of states and the rapid change of the group velocity around the band edge, the spectral conductivity of the pudding mold type band structures becomes larger than that of the usual parabolic band structures. It is found that the pudding mold type band has a coexistence of large Seebeck coefficient and large electric conductivity, and small Lorenz number in the Wiedemann--Franz law due to the specific band shape.
We also find that the low dimensionality of the band structure can contribute to large electronic conductivity and hence a small Lorenz number.
We conclude that the pudding mold type band, especially in low dimensional systems, can enhance not only the power factor but also the dimensionless figure of merit due to stronger reduction of the Lorenz number.
\end{abstract}


\maketitle 

\section{Introduction}\label{introduction}
The thermoelectric effect is one of the most important effect that can contribute in harvesting energy by converting heat into electricity. The efficiency of a thermoelectric material is usually evaluated by the dimensionless figure of merit $ZT$ defined as below,
\begin{eqnarray}
  ZT = \frac{\sigma S^2}{\kappa}T, \label{eq_ZT}
\end{eqnarray}
where $\sigma$, $S$, $T$ and $\kappa$ are the electrical conductivity, Seebeck coefficient, 
temperature, and thermal conductivity, respectively.
The thermal conductivity $\kappa$ consists of electronic and 
lattice (phonon) contributions, $\kappa_{\rm e}$ and $\kappa_{\rm lat}$, respectively. Good thermoelectric materials should have a combination of 
large power factor $PF = \sigma S^2$ and low thermal conductivity.
However, the difficulty of obtaining good thermoelectric materials is generally recognized. Namely, there is generally   
an anti-correlation between the Seebeck coefficient and the electrical conductivity, so that the power factor is maximized at a certain carrier doping level\cite{Mahan}. Moreover,  there is a relation between the electronic thermal conductivity $\kappa_{\rm e}$ and the electrical conductivity $\sigma$, the Wiedemann--Franz law\cite{Wiedemann}, which shows that the ratio between the two is a constant. This implies that it is generally difficult to have high electrical conductivity and low electronic thermal conductivity at the same time.
The lattice thermal conductivity $\kappa_{lat}$ can almost be independent from the electronic structure, so that well known thermoelectric materials have small lattice thermal conductivity and small electronic thermal conductivity, but often moderate power factor.\cite{SnSe}

Theoretically, condition for obtaining high thermoelectric performance has been widely investigated. Most of the previous theories that analyze this condition on a general basis have adopted a parabolic band structure. There, it has been generally recognized that the $B$ factor ($B \sim \mu_e m^{*3/2}/\kappa_{lat}$, where $\mu_e$ and $m^*$ are the mobility and effective mass, respectively), gives a measure for the dimensionless figure of merit.\cite{Chasmar,Mahan2} It has been considered that in order to increase the $B$ factor, a large effective mass and small lattice thermal conductivity are required, but more recently, it has been pointed out that the $B$ factor is actually inversely proportional to the effective mass if the effective mass dependence of the quasi-particle lifetime is taken into account\cite{Yan}. The one dimensional quantum wire and quantum-well structure can obtain large dimensionless figure of merit because of a large $B$ factor.\cite{Hicks1, Hicks2}
Also, it has theoretically been suggested for a coexistence between large 
Seebeck coefficient and large electrical conductivity due to the multi valley effect\cite{Mahan} because several bands can contribute to the electrical conductivity.
From the viewpoint of optimizing 
the power factor, we have proposed that it is better to go beyond the parabolic band structure, and introduced the idea of the ``pudding mold type band'', which has a flat portion at the band edge, gives rise to a coexistence of large Seebeck coefficient and electrical conductivity.\cite{pudding_Kuroki}
We have shown that the shape of the band structure plays an important role for enhancing the thermoelectric performance in a number of materials.\cite{pudding_Kuroki, pudding_Usui}

The dimensionless figure of merit can be also described as follows,
\begin{eqnarray}
  Z_eT &=& \frac{S^2}{\alpha L_0}, \label{eq_ZeT1}\\
  ZT &=& \beta Z_e T, \\
  \beta &=& \frac{\kappa_e}{\kappa_e + \kappa_{lat}}, \label{eq_ZeT3}
\end{eqnarray}
where $L_0 = 2.44 \times 10^{-8}{\rm W\Omega K^{-2}}$ is the Lorenz number in the Wiedemann--Franz law in the degenerate limit the low temperature regime, 
	and the Lorenz number can be described as $L = \alpha L_0$, where $\alpha$ shows the difference between the actual Lorenz number and that in the degenerate limit.

	$Z_e T$ is the dimensionless figure of merit at $\kappa_{lat}/\kappa_e = 0$.
The coefficient $\alpha$ plays a key role in Eq.(\ref{eq_ZeT1}) because the small $\alpha$ enhances $Z_e T$ and hence $ZT$.
It suggests that the Lorenz number in the low carrier concentration regime strongly controls the dimensionless figure of merit.
The Lorenz number should depend on the electronic band structure, so that 
we can assume that the coefficient $\beta$ can be controlled by the shape of the band structure as is the case with the enhancement of the power factor in the pudding mold type band.
Actually, the theoretical result has shown that the anomalous thermal conductivity can be understood from the band structure point of view.\cite{Takeuchi1,Takeuchi2}

Given this background, we study the correlation between the thermoelectric properties and the band structure using Boltzmann equation in the $d\ (= 1, 2 \ {\rm and}\ 3)$ dimensional system.
In this paper, we also study the band structure dependence of the Lorenz number.
We adopt the electronic band dispersion $E({\bf k}) \sim |{\bf k}|^n$ with $n=2,4,6$, and focus on the effect of the band shape, while the effect of the band width on thermoelectric properties is eliminated by normalizing the width.
The pudding mold type band, which has large density of states at the band edge, is obtained for $n \geq 4$ for $d = 2$ and $3$, and $n \geq 2$ for $d=1$.
We find that the pudding mold type band suppress the Lorenz constant compared to that of the usual parabolic band for $d = 3$ due to its shape.
It is because of the band shape and the difference of the energy range, which contributes to the electrical conductivity, Seebeck coefficient and electronic thermal conductivity.
The dimensionless figure of merit is larger in the pudding mold type band structure than that in the parabolic type band due to the large power factor and small Lorenz number.
We conclude that the pudding mold type band can strongly enhance not only the power factor but also the dimensionless figure of merit due to stronger reduction of the Lorenz number.

\section{Method}
In order to understand the band structure dependence of thermoelectric properties, we calculate the thermoelectric properties using the band structures described as follows,
\begin{eqnarray}
  E_n({\bf k}) = A_n|{\bf k}|^n, \label{eq_band}
\end{eqnarray}
where ${\bf k}$ is the wave vector, and $n=2,4,6$ is an index that determines the band shape. $A_n$ is the normalization factor of the band width.
The normalization factor $A_n$ is defined as $A_n|{\bf k}|^n = 1$ at the Brillouin zone edge $(\pi/a,\pi/a,\pi/a)$, and hence the band width is the unit of the energy. We assume a simple cubic unit cell and take the lattice constant $a=1$ hereafter.

Using the Boltzmann equation, the Seebeck coefficient, electrical conductivity and electronic contribution to the thermal conductivity are described as follows,
\begin{eqnarray}
  K_m &=& \sum_{\bf k} \tau({\bf k})v_x^2({\bf k})\left( -\frac{df}{d\varepsilon}\right) (\varepsilon_{\bf k} - \mu)^m, \label{eq_Km}\\
  \sigma &=& e^2 K_0, \label{eq_sigma}\\
  S &=& \frac{1}{eT}\frac{K_1}{K_0}, \\
  \kappa_{e} &=& \frac{1}{T} \left(K_2 - \frac{K_1^2}{K_0} \right), \label{eq_boltzmann}
\end{eqnarray}
where $\tau({\bf k}), v_x({\bf k}), \varepsilon_{\bf k}, f$ and $\mu$ are the wave vector, the quasi-particle lifetime,
the $x$ component of the group velocity $v_x = \hbar^{-1} \partial \varepsilon_{\bf k}/\partial k_x$, the electronic band dispersion, Fermi--Dirac distribution function, and the chemical potential, respectively. We will take $\hbar = 1$ in this article.
The band structure in Eq.(\ref{eq_band}) is isotropic with respect to the wave vector (only depends on $|{\bf k}|$), so that we only show the $xx$ component of the thermoelectric properties.
We approximate the quasi-particle lifetime $\tau$ as a constant ($\tau = 1$).
Theoretical results have shown that the lifetime constant approximation has well explained the experimental results\cite{pudding_Kuroki,pudding_Usui}.
The Seebeck coefficient does not depend on the quasi-particle lifetime within the lifetime constant approximation but the electrical and electrical thermal conductivity depends on the lifetime.

When we assume $\mu/k_BT \gg 1$, a Sommerfeld expansion can be performed in Eq.(\ref{eq_Km}), (\ref{eq_sigma}) and (\ref{eq_boltzmann}),
\begin{eqnarray}
  \sigma(E) &=& \sum_{\bf k} \tau({\bf k}) v^2({\bf k}) \delta(E - E({\bf k})), \\
  \sigma &=& e^2 \sigma(\mu), \label{eq_sommer_sigma}\\
  \kappa_e &=& \frac{\pi^2}{3}k_B^2T\sigma(\mu), \label{eq_sommer_kappa}
\end{eqnarray}
where $\sigma(E)$ and $k_B$ are the spectral conductivity and the Boltzmann constant, respectively. 
The Lorenz number $L_0$ is obtained from Eqs.(\ref{eq_sommer_sigma}) and (\ref{eq_sommer_kappa}),
\begin{eqnarray}
  \frac{\kappa_e}{\sigma T} &=& \frac{\pi^2k_B^2}{3e^2} = L_0.
\end{eqnarray}

Before closing this section, we comment on the validity of the present approach.
We have checked the validity by comparing it with the linear response theory. 
When we assume a realistic quasi particle lifetime (e.g. $\tau = 2.2 \times 10^{-14}$s evaluated in Bi$_2$Te$_3$\cite{Luo}), 
the thermoelectric properties around $\mu \sim 0$, where the power factor is maximized within the Boltzmann theory roughly reproduces those within the linear response theory under constant $\tau$ at room temperature.
We have also checked that the tendency of the band structure dependence of the thermoelectric properties obtained in the following is essentially the same between the Boltzmann theory and the linear response theory. 

\section{Results and Discussion}\label{results}
\subsection{Band structure and spectral conductivity}\label{band}

\begin{figure}[t]
         \includegraphics[width=7.0cm]{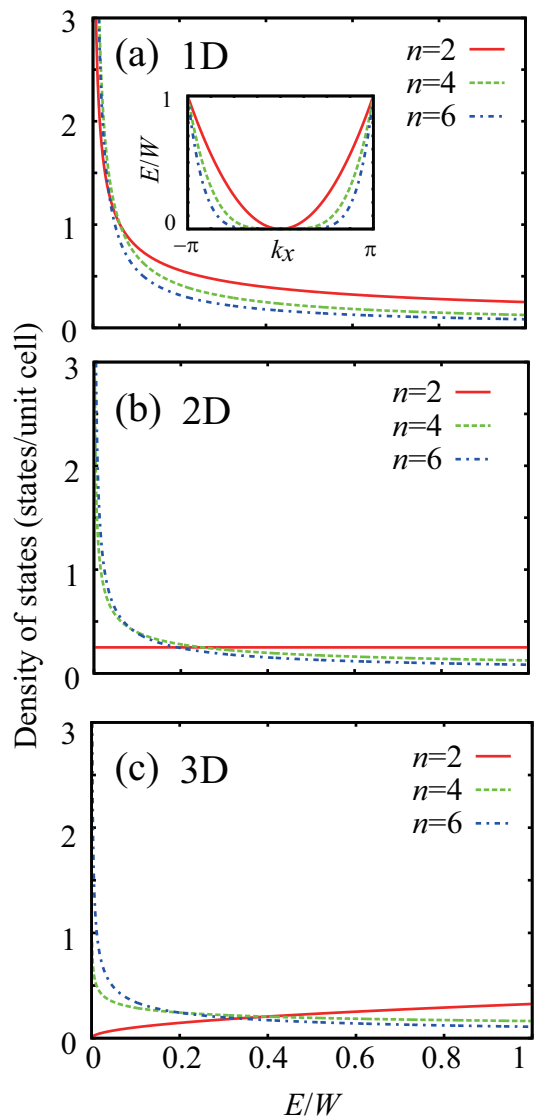}%
          \caption{The density of states for (a) one-, (b) two-, and (c) three-dimensional systems. The inset shows the band structure in the one dimensional case\label{fig1}}%
\end{figure}

We now move on to the results and discussion.
The density of states $D(E)$ for the $d$ dimensional system,
\begin{eqnarray}
	D(E) = c_d A_n^{-d/n}\frac{E^{(d-n)/n}}{n}, \label{eq_dos}
\end{eqnarray}
is shown in Fig.\ref{fig1}, where $c_d$ is a constant parameter that depends on the dimensionality of the system, i.e., $c_1 = 1/\pi, c_2 = 1/4\pi, c_3 = 1/6\pi^2$. 
It can be seen in Fig.\ref{fig1} that the density of states diverges at the edge of the band structure for $n \geq 4$.
This is because the band structure for $n \geq 4$ has a flat portion at the band edge (see the inset of Fig.\ref{fig1}(a)).
We hence call the band structure for $n \geq 4$ the pudding mold type band.
The density of states at the band edge in the one dimensional system diverges for $n \geq 2$ due to a specific feature of the one dimensional system.
The band structure in the one dimensional system also has a flat portion along the $k_y$ and $k_z$ directions.
Therefore, we can regard the band structure in the one dimensional system as the pudding mold type band in the wide sense of the term.

 \begin{figure}[htbp]
        \includegraphics[width=7.0cm]{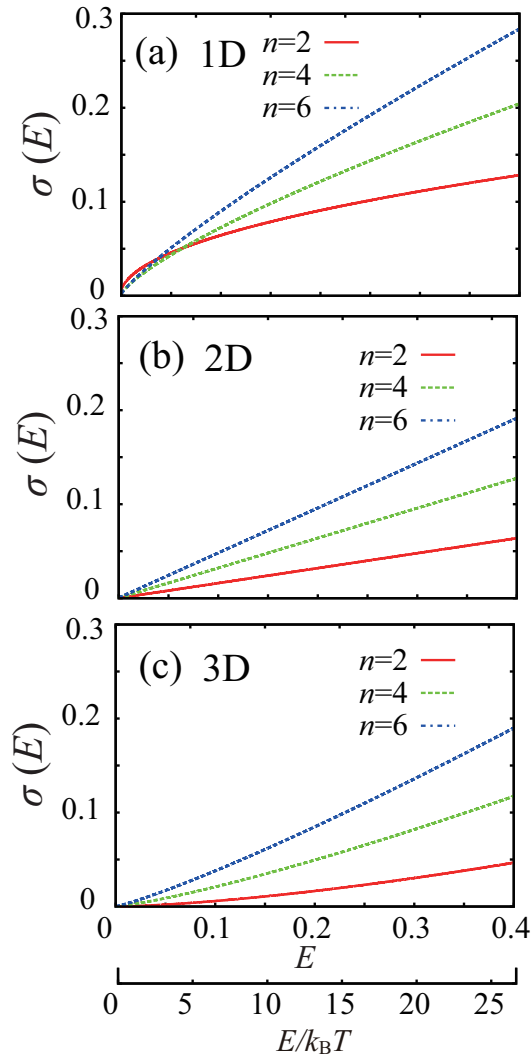}%
         \caption{The spectral conductivity for $d= (a)1, (b)2$ and (c)3.\label{fig2}}%
 \end{figure}

The electronic contribution to the thermoelectric properties only depends on the spectral conductivity $\sigma(E)$. Thus, we now discuss the difference of the spectral conductivity among the band structures before we show the results of the Seebeck effect.
When we define the spectral conductivity $\sigma(E)$ as described below,
\begin{eqnarray}
  \sigma(E) = v^2(E)D(E), \label{eq_sigma_e}
\end{eqnarray}
The group velocity $v(E)$ as a function of energy can be described as follows,
\begin{eqnarray}
  v(E) = A_n^{1/n} n E^{(n-1)/n}. \label{eq_v}
\end{eqnarray}
The spectral conductivity shown in Fig.\ref{fig2} is therefore obtained from Eqs.(\ref{eq_dos})-(\ref{eq_v}),
\begin{eqnarray}
  \sigma(E) = c_d A_n^{(2-d)/n} nE^{(n+d-2)/n}.
\end{eqnarray}
Eq.(\ref{eq_sigma_e}) shows that the contribution of the group velocity to the spectral conductivity is larger than that of the density of states.
This is because, roughly speaking, $D(E) \sim 1/v(E)$ holds.
The spectral conductivity in two and three dimensional systems (Fig.\ref{fig2} (b) and (c)) shows that the spectral conductivity of the pudding mold type band is larger than that for $n=2$. The pudding mold type band has  small group velocity at the band edge, but large velocity right above it, so 
combining the large density of states around the band edge along with the strong increase of the velocity with increasing energy, the pudding mold type band gives rise to a large spectral conductivity. 
In the case of the one dimensional system, the pudding mold type band is obtained for $n \geq 2$, so that all three have the same tendency of the spectral conductivity around the band edge. 

\subsection{Thermoelectric properties of the  3D system}\label{3D}

 \begin{figure*}[htbp]
        \includegraphics[width=17.0cm]{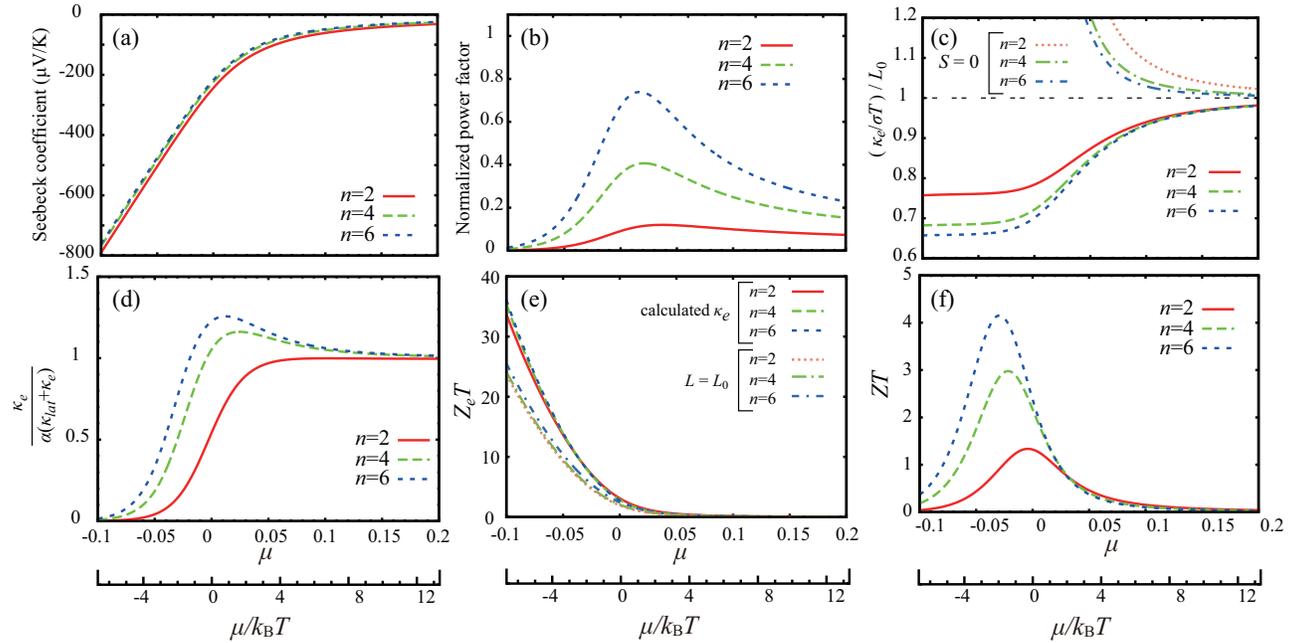}%
	\caption{Calculated thermoelectric properties as functions of the chemical potential. (a) The Seebeck coefficient, (b) the normalized power factor, (c) the coefficient $\alpha$, (d) the coefficient $\gamma$, (e) the dimensionless figure of merit when $\kappa_{lat}/\kappa_e = 0$, and (f) the dimensionless figure of merit for finite $\kappa_{lat}$.\label{fig3}}%
 \end{figure*}

We now present the calculation results of the thermoelectric properties at the temperature $T = 0.015$ throughout the paper, which corresponds to about 350K assuming the band width (the unit of the energy) to be 2eV. We first start with 3D.  The Seebeck coefficient and normalized power factor are shown in Fig.\ref{fig3}(a) and (b), respectively, as functions of the chemical potential. The Seebeck coefficient is barely dependent on the band shape, while the power factor is strongly enhanced for the pudding mold type bands.

 \begin{figure}[htbp]
         \includegraphics[width=7.0cm]{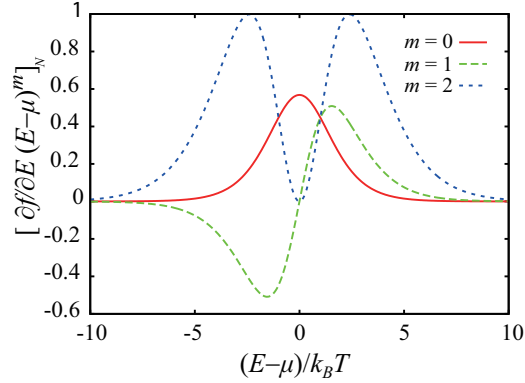}%
	 \caption{Normalized $\frac{\partial f}{\partial E}(E-\mu)^m$, which depicts the energy range that contributes to $K_0$ ($m=0$), $K_1$ ($m=1$) and $K_2$ ($m=2$). \label{fig4}}%
 \end{figure}

In order to understand these band shape dependences, we discuss the energy range of the states that  mainly contribute to the the thermoelectric properties.
$K_m$ in Eq.(\ref{eq_Km}) can be described using the spectral conductivity,
\begin{eqnarray}
  K_m = \int \sigma(E) \left(\frac{\partial f}{\partial E}\right)(E - \mu)^m dE.
\end{eqnarray}
Fig.\ref{fig4} shows $\frac{\partial f}{\partial E}(E-\mu)^m$, which shows the energy range of the states that contribute to the electrical conductivity ($K_0$), Seebeck coefficient ($K_0$ and $K_1$) and electronic thermal conductivity (mainly $K_2$ in the metallic regime).
This figure clearly shows that the contribution to the electrical conductivity comes from around the chemical potential (to $2k_BT$), while the states in the energy range between $k_BT$ and 4$k_BT$ measured from the chemical potential contribute to $K_1$ for the Seebeck coefficient. For the electronic thermal conductivity, $\frac{\partial f}{\partial E}(E-\mu)^2$ mainly takes large values between $E-\mu=k_BT$ and $5k_BT$.

From the above observation, it can be seen that the main contribution to the Seebeck coefficient  comes from $\sigma(\mu)$ for $K_0$, and $\sigma(\mu+2k_BT)$ for $K_1$.
In general, the Seebeck coefficient tends to become large as the chemical potential becomes smaller and sinks below the band edge because $K_0$ decreases due to small $\sigma(\mu)$ while $K_1$ keeps its value because $\sigma(\mu+2k_BT)$ is still large compared to $\sigma(\mu)$.
Now, the band shape dependence of the electrical conductivity and Seebeck coefficient can also be understood within the framework of the energy dependence of $\sigma(E)$.
The spectral conductivity $\sigma(E)$ for $n = 6$ is large compared to that for $n = 2$, so that the electrical conductivity for $n = 6$ is larger than that for $n = 2$ for the same chemical potential.
On the other hand, the Seebeck coefficient does not strongly depend on the band shape because not the value of $\sigma(E)$ itself but the ratio of $\sigma(E)$ between $E \sim 0$ and $E \sim 4 k_BT$ mainly determines the Seebeck coefficient.
Therefore, we can say that the pudding mold type band gives rise to a large power factor compared to the parabolic band, for {\it a fixed chemical potential}, because the electrical conductivity is large while the Seebeck coefficient is essentially band shape independent.

\begin{figure}[htbp]
	\includegraphics[width=6.0cm]{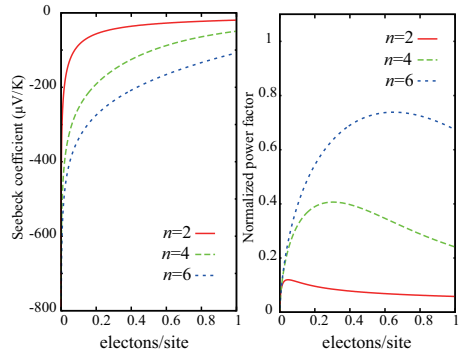}
	\caption{(a) The Seebeck coefficient and (b) power factor as a function of the carrier doping ratio. for the three dimensional band structures.\label{fig3-2}}
\end{figure}

Here we note the difference between fixed chemical potential and fixed carrier concentration. To see this, we show in Fig.\ref{fig3-2}, the Seebeck coefficient and the normalized power factor as functions of the number of electrons per unit cell. As was shown in ref.\onlinecite{pudding_Kuroki}, the power factor is optimized at a much larger carrier concentration in the pudding mold type band compared to the parabolic band.
In the preceding context, this is because the amount of electrons is much larger in the former compared to the latter at a fixed chemical potential.

We now turn to the electronic thermal conductivity and the Wiedemann--Franz law. We show  in Fig.\ref{fig3}(c) the coefficient $\alpha = \kappa_e/(\sigma T)/L$ in Eq.(\ref{eq_ZeT1}) here again as a function of the chemical potential.  $\alpha$ should be unity when the Lorenz number in the degenerate limit is valid.
It is found that the Lorenz number almost equals to $L_0$ in the large chemical potential regime ($\mu > 5 k_BT$) because the electronic structure within $\pm 5k_BT$ measured from the chemical potential mainly contributes to $K_m$.
When the chemical potential decreases from $\mu \sim 5k_BT$, $\alpha$ decreases in all of the band structures, but the reduction of the Lorenz number for $n = 4$ and $6$ is larger than that for $n=2$.
In order to understand the origin of the band shape dependence of $\alpha$, we describe $\kappa_e/(\sigma T)$ as follows,
\begin{eqnarray}
  \frac{\kappa_e}{\sigma T} = \frac{1}{e^2T^2} \frac{K_2}{K_0} - S^2. \label{eq_Wiedemann}
\end{eqnarray}
When we assume $\mu \gg k_BT$, the second term of the right hand side in Eq.(\ref{eq_Wiedemann}) should be zero, and $K_2$ and $K_0$ can be described using the Sommerfeld expansion, and we obtain $\kappa_e/\sigma T = L$.
Eq.(\ref{eq_Wiedemann}) shows that large Seebeck coefficient always suppresses $\alpha$, while $\alpha$ tends to be enhanced by the large $K_2/K_0$ in the low carrier concentration regime as can be understood from a discussion similar to that of the chemical potential dependence of the Seebeck coefficient.
The dotted line in Fig.\ref{fig3}(c) shows the first term of the right side in Eq.(\ref{eq_Wiedemann}) and the value monotonically increases when the chemical potential decreases.
As we described in the case of the Seebeck coefficient, $K_2/K_0$ roughly corresponds to $\sigma(\mu+4k_BT)/\sigma(\mu)$.
If we compare $\sigma(E)$ between $n = 2$ and $6$, $\sigma(E) \sim E^{2/3}\ (n=2)$  and $\sigma(E) \sim E\ (n = 6)$, so that $\sigma(\mu+4k_BT)/\sigma(\mu)$ for $n = 2$ is larger than that for $n = 6$.
Therefore, Comparing the calculated Lorenz number $L$ with $L_0$, the Lorenz number is more strongly reduced for the pudding mold type band. This is also reflected in $Z_e T$ shown in 
Fig.\ref{fig3}(e), where adopting the actual $\alpha$ values results in about 30\% increase in $Z_eT$  in the small chemical potential regime compared to the case when we assume that the Wiedemann--Franz law is valid ($\alpha = 1$).

Since $Z_eT$ is the theoretically allowed maximum dimensionless figure of merit, Fig.\ref{fig3}(e) indicates that wide-band-gap materials (where the chemical potential can be lowered sufficiently below the band edge without encountering the lower bands) has a potential for extremely large dimensionless figure of merit, regardless of the reduction of the Lorenz number. Also, $Z_eT$ calculated as functions of the chemical potential is almost independent of the band shape. These result are obtained because, in the absence of the lattice thermal conductivity, the dimensionless figure of merit in the low carrier limit is almost solely determined by the Seebeck coefficient owing to $\sigma T/\kappa_e\sim $ constant in this regime. In reality, actual materials have finite lattice thermal conductivity, so that $\kappa_{lat}/\kappa_e$ is large in the low carrier concentration regime.
Therefore, the ``ideal" dimensionless figure of merit $Z_e T$ for $\kappa_{lat} = 0$ never corresponds to the ``actual" dimensionless figure of merit.
In fact, the actual dimensionless figure of merit $ZT$ depends on the shape of the band structure.
In Fig.\ref{fig3}(f), we show the dimensionless figure of merit $ZT$ calculated by  assuming the lattice thermal conductivity $\kappa_{lat}$ to be the same as $\kappa_e$ for the electron doping level of $n_e = 0.03$ electrons/site for $n = 6$ in the one dimensional system, which is a reasonable choice considering values of the thermal conductivity in actual materials.
In contrast to the case of $Z_e T$, the effect of the band shape on $ZT$ is clearly seen, where the pudding mold type band (the case of $n=6$) indeed strongly enhances the thermoelectric efficiency.
In order to pin down the origin of the enhancement of $ZT$ for the pudding mold type band, we introduce $\gamma = \kappa_e/\alpha(\kappa_e + \kappa_{lat})$, which is a part of $ZT$, namely,
\begin{eqnarray}
	ZT = \frac{\sigma S^2}{\kappa_e + \kappa_{lat}} = \frac{S^2}{L}\frac{\kappa_e}{\alpha(\kappa_e + \kappa_{lat})} = \gamma \frac{S^2}{L}.
\end{eqnarray}
$\gamma$ is thus a coefficient that determines to what extent the dimensionless figure of merit deviates from its ``ideal" value with $\alpha=1$ and $\kappa_{lat}=0$.
The pudding mold type band, the combination of small $\alpha$ and large $\kappa_e/\kappa_{lat}$ result in the enhancement of the dimensionless figure of merit as compared to that in the parabolic band result. Physically, the enhancement of $ZT$ in the pudding mold band should be attributed to the large conductivity $\sigma$, which is almost proportional to the electronic thermal conductivity $\kappa_e$ in the low carrier density regime for a fixed temperature as seen in Fig.\ref{fig3}(d). 
Indeed, combining Figs.\ref{fig3}(a) (the Seebeck coefficient is essentially independent of the band shape for a fixed chemical potential) and (b) (large power factor for $n=4$ and $n=6$) shows that the pudding mold type band has large electrical conductivity when compared at the same chemical potential. Therefore, the ratio $\kappa_e/\kappa_{lat}$ is larger for the pudding mold type band than that for the parabolic band.
Note that for some range of chemical potential, $\gamma>1$ for the pudding mold band, implying that the dimensionless figure of merit with finite $\kappa_{lat}$ exceeds the ``ideal'' $ZT$ value of $\alpha=1$ and $\kappa_{lat}=0$ due to $\alpha<1$.
We summarize that the pudding mold type band has a large power factor and large dimensionless figure of merit compared to the parabolic band.

We stress that having a flat band bottom as in the pudding mold type band is different from simply increasing the mass in a parabolic band. In fact, it has recently been realized that enhancing the mass in a parabolic band actually is {\it unfavorable} for good thermoelectric properties. Namely, when the electron-phonon scattering is dominant in the electron transport process, the quasi-particle lifetime $\tau$ is proportional to $m^{*-3/2}$.\cite{Yan} Taking this effect into account,  the $B$ factor is actually inversely proportional to the effective mass. In the pudding mold type band we considered here, the effective mass at the band bottom, namely, the inverse of the coefficient of the term proportional to $|{\bf k}|^2$, diverges, but this does not imply that the lifetime is 0 because the group velocity rapidly increases just above the band edge (note that the band width is kept to be the same as that of the parabolic band, which requires a rapid increase of the band dispersion just above the band edge). Hence, the pudding mold type band has an ideal band shape in the sense that the large group velocity just above the Fermi level, combined with  the large carrier concentration due to the large DOS at the band edge, gives rise to excellent thermoelectric properties compared to the parabolic band.

\subsection{The thermoelectric properties of  2D and 1D systems}\label{2D}

 \begin{figure}[htbp]
        \includegraphics[width=7.0cm]{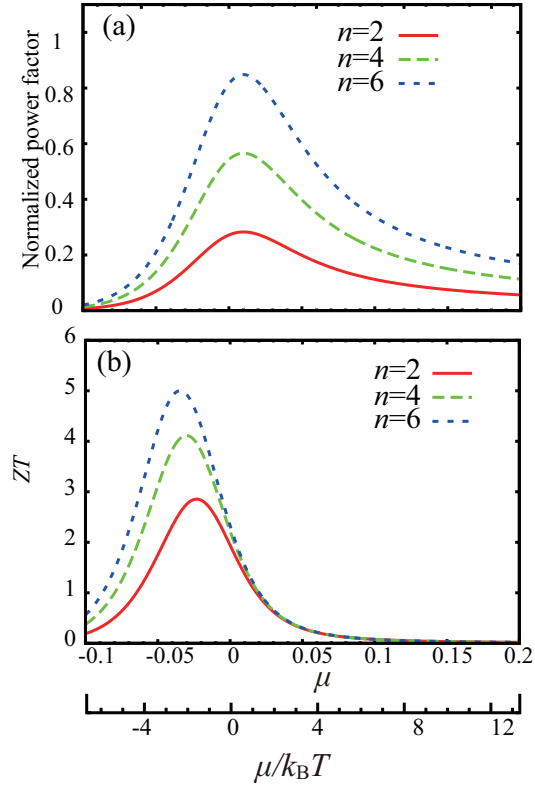}%
         \caption{(a) The normalized power factor and (b) the dimensionless figure of merit for the two dimensional band structure.\label{fig5}}%
 \end{figure}

We now move onto the results of the thermoelectric properties in the one and two dimensional band structures.
The spectral conductivity $\sigma(E)$ for the 2D case is proportional to the energy as far as the band dispersion is isotropic (dependent only on $|{\bf k}|$) as in Eq.(\ref{eq_band}).
Therefore, the Seebeck coefficient ($\sim K_0/K_1$) and $\kappa_e/(\sigma T)$ (consisting of $K_2/K_0$ and $S$) do not depend on the band shape because these quantities are determined by ratios of $\sigma(E)$ at different energy range. 
On the other hand, the power factor and the dimensionless figure of merit shown in Fig.\ref{fig5} indicates the importance of the band shape.
It is because the absolute value of the spectral conductivity in Fig.\ref{fig2}(b) is larger for the pudding mold type band due to the large density of states and large difference of the group velocity around the edge of the band structure.
Therefore, the effect of the pudding mold type band does not strongly depend on the dimensionality of the band structure.

If we compare the pudding mold type band between the three and two dimensional systems, the spectral conductivity $\sigma(E)$ is (almost) proportional to the energy in both cases.
This means that the Seebeck coefficient and the coefficient $\alpha$ in Eq.(\ref{eq_ZeT1}), which measures the deviation from the Lorenz number $L_0$, are almost independent of the dimensionality of the system, so that the dimensionless figure of merit also does not strongly depend on the dimensionality. This is in contrast to the case of the parabolic band, where 2D has much larger $ZT$ than in 3D (compare Fig.\ref{fig3}(f) and Fig.\ref{fig5}(b)).
The density of states of the parabolic band drastically changes from $\sqrt{E}$ to constant, so that the absolute value of the spectral conductivity for 2D case is larger than that for 3D case because the number of wave vector for contributing the conduction along the same axis increases due to the reduction of the dimensionality.
Thus the power factor is enhanced in the low dimensional system, so that the dimensionless figure of merit increases.

 \begin{figure}[htbp]
        \includegraphics[width=7.0cm]{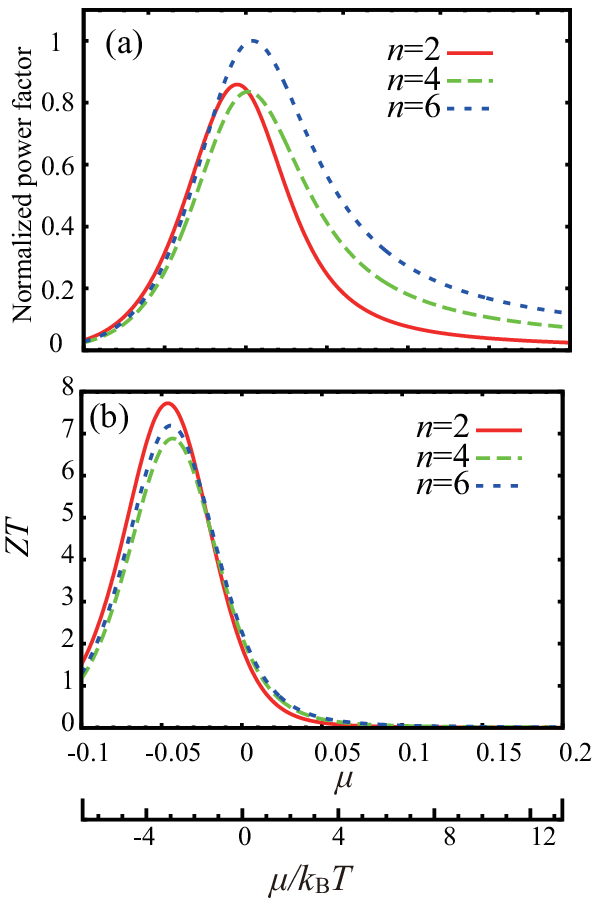}%
         \caption{(a) The normalized power factor and (b) dimensionless figure of merit for the one dimensional band structure.\label{fig6}}%
 \end{figure}

The thermoelectric properties of the one dimensional system is shown in Fig.\ref{fig6}.
It is found that the power factor and the dimensionless figure of merit does not depend on $n$ because all of the band structures belong to the pudding mold type band.
Moreover, the dimensionless figure of merit (calculated for finite $\kappa_{lat}$) reaches about 7, which results from the combination of large power factor and small $\kappa_e/(\sigma T)$ .
The dimensionless figure of merit has the largest maximum value for $n=2$ because in the case of 1D, the $\sigma(E)$ is strictly proportional to the velocity $v(E)$, which has the largest values for $n=2$ in the low energy regime.

In the one dimensional case, the thermoelectric properties does not strongly depend on the band structures for the temperature/band width adopted in the present paper.
However, as seen in Fig.2(a), $\sigma(E)$ is larger for $n = 4, 6$ than for $n = 2$ (the parabolic band) at higher energy, which means that the power factor is larger in the former than in the latter if the temperature is raised (or the band width is reduced). This difference in $\sigma(E)$ originates from the  group velocity of the dispersive portion of the band structure, namely, band structures with larger $n$ have steeper gradient away from the band edge. This indicates that not only the divergence of the density of states at the band edge, but also the dispersiveness of the band structure away from the edge is important to have enhanced thermoelectric properties. 

\subsection{Comparison to the tight binding model}\label{tight}

We now discuss the effect of the band shape on the thermoelectric properties using more realistic tight binding models in the two dimensional case. We shall see below that the above conclusions do not qualitatively change even for realistic band structures.
The band structures are described as follows,
\begin{eqnarray}
  E({\bf k}) = -2t ({\rm cos}k_x + {\rm cos}k_y) - 4t'{\rm cos}k_x{\rm cos}k_y,
\end{eqnarray}
where $t$ and $t'$ is the nearest and next nearest hopping integral in the square lattice.
The next nearest hopping integral controls the band shape and we take $t = 0.125, t'=0, -0.45t$ and $-0.5t$. The band width is kept at unity for these choices.
In the band structure of $t'=-0.5t$, the density of states at the band edge diverges due to the van-Hove singularity.
Similary, the one dimensional tight binding model is given as, 
\begin{eqnarray}
  E(k) = -2t {\rm cos}k -2t' {\rm cos}2k,
\end{eqnarray}
where we take $t=0.25$ and $t' = 0, -0.25$.
The shape of the band structure becomes like a pudding mold for $t'=-0.25$.

 \begin{figure}[htbp]
         \includegraphics[width=7.0cm]{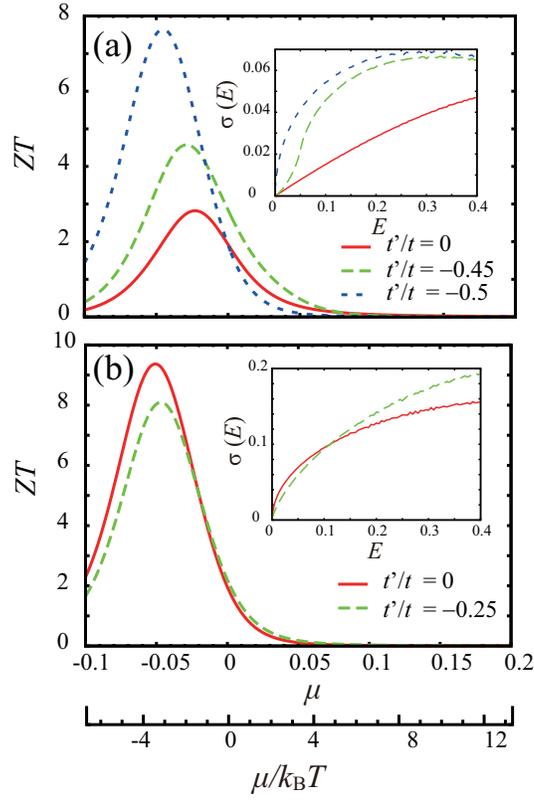}%
          \caption{The dimensionless figure of merit for (a) two- and (b) one-dimensional tight binding models. The inset shows the spectral conductivity.\label{fig7}}%
  \end{figure}

\begin{figure}[httbp]
	\includegraphics[width=6cm]{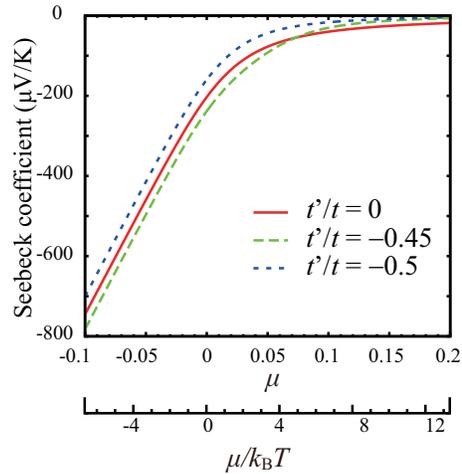} 
	\caption{The Seebeck coefficent of the two-dimensional tight binding models.\label{fig8}}
\end{figure}

Fig.\ref{fig7} shows the dimensionless figure of merit for the one and two dimensional tight binding models.
As for the two dimensional tight binding case, the spectral conductivity for $t'=0$ in Fig.\ref{fig7}(inset) is almost the same as that for the parabolic band for the two dimensional system.
The spectral conductivity for $t'=-0.5$ is larger than that for $t'=0$ due to the shape of the band structure, while the energy dependence of $\sigma(E)$ itself is different from that of isotropic pudding mold type band (Eq.(\ref{eq_sigma_e}) given in Fig.\ref{fig2}.
This is because the band dispersion of the tight binding model strongly depends on the direction of the wave vector. Nonetheless, the conclusions drawn in the former part of the paper remains completely valid.
As shown in Fig.\ref{fig8}, despite the quite different behavior $\sigma(E)$ compared to that of the simplified band structure, $S(\mu)$ is not strongly band shape dependent even for the tightbinding model.
The dimensionless figure of merit is larger for the pudding mold type band ($t'=-0.5$) than that for the parabolic band ($t'=0$) consistent with the discussion in Secs.\ref{3D} and \ref{2D}.
In the one dimensional case, the spectral conductivity and the dimensionless figure of merit is almost the same for $t'=0$ and $-0.25$, which is also consistent from the discussion of Sec.\ref{2D}.
In total, we can conclude  that the pudding mold type band of the tight binding model can give rist to high thermoelectric performance, as in the case of isotropic band structures.


\section{Conclusion}\label{conclusion}

In order to investigate the effect of the shape of the band structure on the thermoelectric properties,
we have calculated the Seebeck coefficient, electrical conductivity, electrical thermal conductivity and dimensionless figure of merit using Boltzmann equation with the band structures defined as $E({\bf k}) = A_n|{\bf k}|^n$ in the $d$-dimensional system ($d = 1,2,3$).
For $n = 2$, we obtain usual parabolic band structures while for $n>3$, the pudding mold type band, which has a flat portion at the band edge, is obtained.
The specific feature of the pudding mold type band is the divergence of the density of states and the rapid increase of the group velocity just above the band edge.
Due to the specific shape of the pudding mold type band, the spectral conductivity becomes larger than that for the parabolic shape.
This gives rise to large power factor and small Lorenz number in the Wiedemann--Franz law.
The dimensionless figure of merit for the pudding mold type band is therefore enhanced due to the specific shape of the band structure.
Moreover, the band structure in the one dimensional system enhances the thermoelectric properties regardless of the band shape because there is always large density of states around the band edge.
We can thus regard any band structure for one dimensional system as the pudding mold type band.
We also calculate the Seebeck coefficient within the tight binding model.
We find that the thermoelectric properties in the pudding mold type band structure does not depend on the calculation models.
In conclusion, the pudding mold type band is efficient for enhancing the dimensionless figure of merit.

We note that the bulk crystal structure does not always have the three dimensional band structure.\cite{pudding_Usui,He}
For example, FeAs$_2$ crystallizes in an orthorhombic Marcasite structure, but the conduction band has low dimensional character due to the anisotropy of the orbitals that are the origin of the bands.\cite{pudding_Usui}
Therefore, by combining the orbital character with the crystal structure, we may be able to find new good thermoelectric materials that have low dimensional band structures. The present study shows that the ``band structure engineering'' is an efficient way of searching new thermoelectric materials.


\begin{acknowledgments}
H.U. is grateful to T. Takeuchi for fruitful discussion. 
This study has been supported by JST-CREST (No. JPMJCR16Q6) and by Grant-in Aid for Scientific Reserach from the Japan Society for the Promotion of Science (No.20110007).
\end{acknowledgments}


\begin{thebibliography}{99}
\bibitem{Mahan}
For a general review on the theroretical aspects as well as 
experimental observations of thermoelectric effect, see, 
G.D. Mahan Solid State Physics {\bf 51}, 81 (1997).
\bibitem{Wiedemann}
R. Franz and G. Wiedemann, Ann. Phys. {\bf 165}, 497 (1853).
\bibitem{SnSe}
L.-D. Zhao, S.-H. Lo, Y. Zhang, H. Sun, G. Tan, C. Uher, C. Wolverton, V. P. Dravid and M. G. Kanatzidis, Nature {\bf 508}, 373 (2014).
\bibitem{Chasmar}
R.P. Chasmar and R. Stratton, J. Electron. Control 7, 52 (1959).
\bibitem{Mahan2}
G. D. Mahan, J. Appl. Phys. {\bf 65}, 1578 (1989).
\bibitem{Yan}
J. Yan, P. Gorai, B. Ortiz, S. Miller, S. A. Barnett, T. Mason, V. Stevanovi$\acute{{\rm c}}$, and E. S. Toberer, Energy Environ. Sci. {\bf 8}, 983 (2015).
\bibitem{Hicks1}
L. D. Hicks and M. S. Dresselhaus, Phys Rev. B {\bf 47}, 12727 (1993).
\bibitem{Hicks2}
L. D. Hicks and M. S. Dresselhaus, Phys. Rev. B {\bf 47}, 16631(R) (1993).
\bibitem{pudding_Kuroki} 
K. Kuroki and R. Arita, J. Phys. Soc. Jpn. {\bf 76}, 083707 (2007).
\bibitem{pudding_Usui}
H. Usui, K. Suzuki, K. Kuroki, S. Nakano, K. Kudo and M. Nohara, Phys. Rev. B {\bf 88}, 075140 (2013).
\bibitem{Takeuchi1}
T. Takeuchi, Zeitschrift fuer Kristallographie {\bf 224}, 35 (2009).
\bibitem{Takeuchi2}
T. Takeuchi, Journal of Electronic Materials {\bf 38}, 1354 (2009).
\bibitem{Luo}
X. Luo, M. B. Sullivan and S. Y. Quek, Phys. Rev. B {\bf 86}, 184111 (2012).
\bibitem{He}
J. He, S. Hao, Y. Xia, S. S. Naghavi, V. Ozolins and C. Wolverton, Chem. Mater. {\bf 29}, 2529 (2017).

\end{thebibliography}

\end{document}